# Improving the astrometric performance of VLTI-PRIMA

J. Woillez[*a], R. Abuter[a], L. Andolfato[a], J.-P. Berger[a], H. Bonnet[a], F. Delplancke[a], F. Derie[a],
N. Di Lieto[a], S. Guniat[a], A. Mérand[b], T. Phan Duc[a], C. Schmid[a], N. Schuhler[b],
T. Henning[c], R. Launhardt[c], F. Pepe[d], D. Queloz[d], A. Quirrenbach[e], S. Reffert[e],
J. Sahlmann[f], D. Segransan[d].

[a]European Southern Observatory (ESO),
Karl-Schwartzschild-Str. 2, Garching bei München 85748, Germany;
[b]European Southern Observatory (ESO),
Alonso de Córdova 3107, Vitacura, Casilla 19001, Santiago de Chile, Chile;
[c]Max-Planck Institut für Astronomie (MPIA),
Königstuhl 17, D-69117 Heidelberg, Germany;
[d]Geneva Observatory,
University of Geneva, 51 ch des Maillettes, 1290 Sauverny, Switzerland;
[e]University of Heidelberg,
Landessternwarte, Königstuhl 12, D-69117 Heidelberg, Germany;
[f]European Space Agency, ESAC,
PO box 78, Villanueva de la Cañada, 28691 Madrid, Spain

## ABSTRACT

In the summer of 2011, the first on-sky astrometric commissioning of PRIMA-Astrometry delivered a performance of 3 m″ for a 10 ″ separation on bright objects, orders of magnitude away from its exoplanet requirement of 50 μ″ ~ 20 μ″ on objects as faint as 11 mag ~ 13 mag in K band. This contribution focuses on upgrades and characterizations carried out since then.

The astrometric metrology was extended from the Coudé focus of the Auxillary Telescopes to their secondary mirror, in order to reduce the baseline instabilities and improve the astrometric performance. While carrying out this extension, it was realized that the polarization retardance of the star separator derotator had a major impact on both the astrometric metrology and the fringe sensors. A local compensation of this retardance and the operation on a symmetric baseline allowed a new astrometric commissioning. In October 2013, an improved astrometric performance of 160 μ″ was demonstrated, still short of the requirements. Instabilities in the astrometric baseline still appear to be the dominating factor.

In preparation to a review held in January 2014, a plan was developed to further improve the astrometric and faint target performance of PRIMA Astrometry. On the astrometric aspect, it involved the extension of the internal longitudinal metrology to primary space, the design and implementation of an external baseline metrology, and the development of an astrometric internal fringes mode. On the faint target aspect, investigations of the performance of the fringe sensor units and the development of an AO system (NAOMI) were in the plan. Following this review, ESO decided to take a proposal to the April 2014 STC that PRIMA be cancelled, and that ESO resources be concentrated on ensuring that Gravity and Matisse are a success. This proposal was recommended by the STC in May 2014, and endorsed by ESO.

## 1    INTRODUCTION

Following a 1998 feasibility study[1], the design of star separators, differential delay lines, fringe sensor units, and metrology supporting the astrometric mode of PRIMA started in 2000. Their installation on VLTI happened in the fall of 2008. The commissioning of the full astrometric mode took 14 technical slots over the following three years, and led to an on-sky astrometric demonstration in the fall of 2011. The status of the instrument and the performance achieved at the time are detailed in Sahlmann et al. (2013)[2]. This contribution is in the continuation of Sahlmann et al. (2012)[3], an update of the Extrasolar Planet Search with PRIMA project (ESPRI[4]), and Schmid et al. (2012)[5], an update of the PRIMA-Astrometry instrument itself.

---

[*] E-mail: jwoillez@eso.org

Section 2 presents the modifications made to the metrology of PRIMA to improve the astrometric performance, as well as the impact and correction of major polarization effects. Section 3 details the improved on-sky performance demonstrated in October 2013 in this new metrology and astrometric baseline configuration. Section 4 presents the performance improvement plans that were formulated in preparation of the PRIMA "Gate Review" held in January 2014.

## 2 EXTENDING THE INTERNAL METROLOGY TO M2

Facing an astrometric performance shortage of ~2 orders of magnitude, the project carried out, between March and May 2012, a system-wide analysis of the instrument, in order to make recommendation on how to best recover the astrometric performance. The extension of the internal metrology to the secondary mirror became the top priority, even though this improvement would probably not be sufficient to meet the astrometric requirements.

By the summer of 2012, the first attempt to operate the PRIMA metrology (PRIMET) up to auxiliary telescope secondary mirror (M2) had failed. The role of beam train polarisation in this failure became obvious when the PRIMET flux returned from the M2 retro-reflector was measured fully modulated by rotation of the field derotator in the Star Separator: for some derotator angles, the returned metrology flux had quasi total extinctions. The first half of 2013 was dedicated to solving this polarisation problem and making PRIMET work up to M2.

The polarisation properties of the full beam train were measured. This was done by inserting, in the PRIMET beam, at the fringe sensor unit entrance, a pair of rotating quarter and half wave plates, and measuring the metrology flux returned from various locations in the beam train. The setup for this measurement is illustrated in Figure 1. The results, given in Table 1, show a large retardance (~100 deg) at the level of the derotator, responsible for the total extinction of the returned metrology flux for some derotator angles.

The impact on the Fringe Sensor Units (FSU) of this significant retardance was also studied. Using AT stations G2-J2 located on opposite North/South sides of the delay line (DL) tunnel, the derotators of AT3 and AT4 had to be operated at 90 deg physical offset of each other. Combined with the retardance of the derotator, this generated a differential retardance between the telescopes and therefore a catastrophic impact on the FSU. It was shown, and verified on sky, that for some derotator orientations, the contrast of the fringes at the FSU would drop to zero. This differential retardance also had a large impact on the stability of the FSU calibrations, specifically the phase offsets of the various FSU ABCD outputs. It also explained why for some targets, fringes could not be found for some sky locations. Full fringe contrast could be recovered when the derotators were operated without offsets, but at the expense of the dual-field operation.

To enable PRIMET operation to M2, the derotator retardance of ~100 deg was compensated by the addition of a quarter-wave plate attached to, and therefore co-rotating with, the derotator assembly. In addition, to mitigate the risk of imperfect derotator retardance compensation with a quarter-wave plate, a new South-South baseline was opened, allowing the operation of the derotators with the same angle, therefore making the beam between the two ATs completely symmetrical.

Finally, the M9 dichroic was upgraded to reflect a larger fraction of the 1319 nm metrology beam, recovering part of the transmission losses resulting from the additional reflections through the telescope. In August 2013, PRIMET became fully operational propagating up to M2.

|  | AT3 | | AT4 | |
| --- | --- | --- | --- | --- |
|  | Retardance [deg] | Dichroism [%] | Retardance [deg] | Dichroism [%] |
| **Switchyard** | 30.25/-8.15 | +0.3/+1.1 | 29.70/-8.50 | +0.1/-2.3 |
| **M12 + M16** | 27.70 | +0.1 | 27.90 | -0.5 |
| **M9** | 23.40 | +5.1 | | |
| **Derotator** | 100.70 | -0.1 | | |
| **Coudé (M8 to M4)** | -12.65 | 5.8 | | |
| **M3** | 7.0 | +1.5 | | |

Table 1: Polarization properties at 1319 nm of the VLTI beam train to AT3 and AT4. These properties, retardance and dichroism, are given for each of the following groups: switchyard, M12+M16, M9, derotator, Coudé, and M3. The properties of AT4 have not been measured; they are assumed identical to AT3. For the switchyard, two values are given; they correspond to different inputs to the fringe sensor units. All other optics are shared between the inputs of a given AT. Note: the retardance of the switchyard, M12+M16, and M9 have no impact on the metrology since these optics are aligned with the PRIMET polarization state.

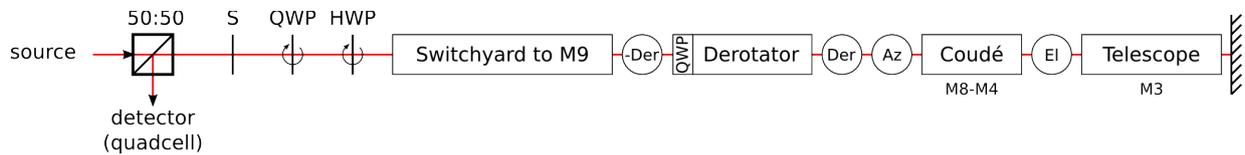

Figure 1: Setup to measure the polarization properties of the VLTI beam train at 1319 nm. Source, detector, 50:50 beam-splitter, and polarizer (S) are realized with PRIMET metrology laser. A quarter-wave-plate and a half-wave-plate are used to modulate the linear polarization state of the laser and explore the polarization properties up to a retro-reflector (hashed vertical line). Derotator, azimuth and elevation rotation angles are shown. These rotation angles allow separating the different beam train sections: switchyard to M9, derotator, Coudé, and telescope.

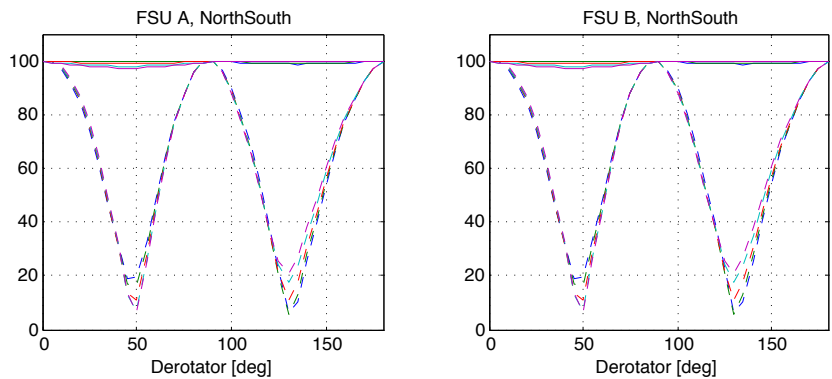

Figure 2: Model of FSU contrast loss as a function of one of the derotator physical angle, for a North-South baseline, based on parameters given in Table 1. Dashed: derotators only. Solid: after derotators retardance compensation with QWPs.

# 3 ASTROMETRIC PERFORMANCE WITH METROLOGY TO M2

In September 2013, with PRIMET operational up to M2, on the new South-South A1-J2 baseline, a new run of 6 nights focused on the astrometric performance of PRIMA. In this section, the new astrometric performance and its limits is presented.

## 3.1 Improved astrometric performance

As expected the performance improved in comparison to the first astrometric commissioning of 2011. Figure 3 shows the astrometric fit results for HD 10360, a bright binary (K~4 for both stars) with 11 arcsec separation. This target had previously been observed[2] at the astrometric commissioning of 2011. As such, it is a good target to compare and assess the improvement brought by the PRIMET extension to M2. The night-to-night stability of the astrometric fits improved by a factor ~19, resulting in ~800 µas peak-to-peak (p-p), compared to the ~15000 µas p-p of 2011. It is worth mentioning that HD 10360, faces the worst observing configuration possible, as the baseline is almost orthogonal to the separation vector. The mean East-West baseline orientation can be guessed from the error ellipses of Figure 3: the minor axis, in the right ascension direction, comes from a baseline oriented in that direction. The HD 10360 pair separation is mostly in the declination direction. The performance evolution summary, given in Figure 4, also gives the extrapolated performance from a 20 m effective baseline to a 100 m North-South baseline. Note that this figure 4 uses peak-to-peak stability for the performance in order to insist on the fact that we have no significant information on the statistics of the biases affecting the measurements.

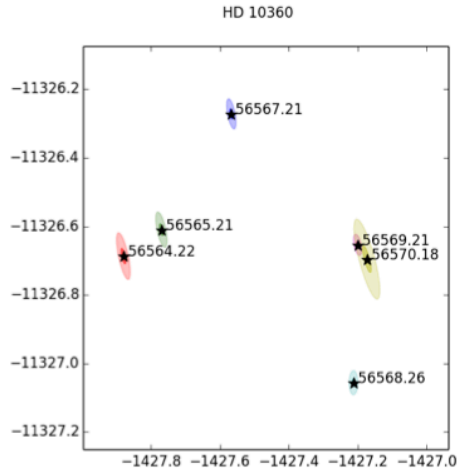

Figure 3: Astrometric fits for 6 nights of HD 10360 observations, in September 2013. The target has an expected motion of 135 µas/night in RA, as illustrated by the hexagons on the right. MJD56564.2 corresponds to 2013-09-28 and so on.

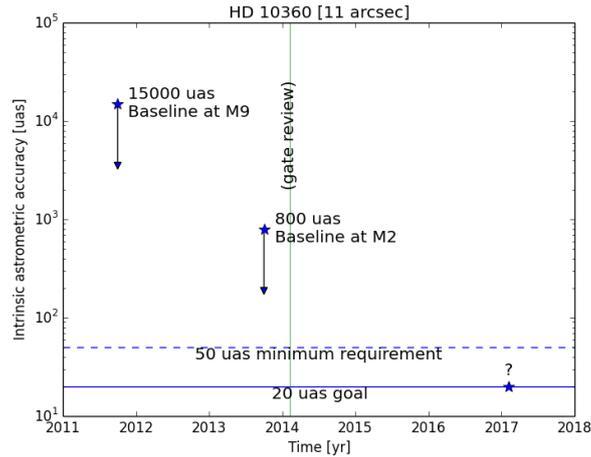

Figure 4: Summary of the performance evolution of PRIMA on bright pairs (estimated from night-to-night peak-to-peak stability), compared to the minimum requirement of 50 μas$_{RMS}$, and the goal of 20 μas$_{RMS}$. The extrapolations (down arrows) correspond to the anticipated opening of an additional North-South baseline that would improve the observing configuration for HD 10360.

### 3.2 Baseline instability investigation

The following baseline instability investigation is based on data collected on the same HD10360 pair.

**Intra night evolution** - The first indication that there is indeed a baseline stability issue comes from the evolution of the astrometric fit within each night. In *Figure 5* - Left, each fit is for a sliding time interval of ~1.5 hrs. In a way, this figure aims at illustrating the evolution of the astrometric fit within each night. The results tend to oscillate with amplitudes up to almost 1500 μas, indicating that the dispersion of 800 μas peak-to-peak might be an underestimate of the underlying baseline instability effect. The shortening of the measurement interval (the time requirement for an astrometric measurement is 1 hour) could also amplify the effect of the baseline instability bias.

**Swaps** - With PRIMET ending at M2, the baseline stability depends on the conjugation of the metrology corner cube to primary space. This conjugation depends on the corner cube itself, the M2 and M1 mirrors, and the pointing of the telescope. The oscillatory bias seen in *Figure 5* - Left could be related to this conjugation operation. Specifically, one possible scenario is a differential telescope pointing between the NORMAL and SWAP states. Consequently, for each night, the NORMAL and SWAP states were fitted independently, producing *Figure 5* - Upper Right. The NORMAL and SWAP solutions do not match. Note also, the change in the error ellipse direction, which could be related to the now dominating uncertainty on the metrology zero point. With this observation, an explanation for the behaviour in *Figure 5* - Left can be offered: within each time interval, the weighting between the NORMAL and SWAP states changes, causing the fit result to oscillate between NORMAL and SWAP biases. This led to the idea of balancing the weighting between the two states. The result of this balancing is shown in Figure 6, where an improvement to 500 μas peak-to-peak is observable.

**Telescope differential pointing** - *Figure 5* – Lower Right shows the differential pointing (Alt and Az) between AT3 and AT4. Besides recording artefacts, differential pointing transitions are related to the SWAP operations and correspond to real astrometric baselines effects. Based on an estimated differential pointing transition of ~0.001 deg for each swap operation, for the altitude and the altitude-corrected azimuth, the baseline motion between the NORMAL and SWAP state can be estimated. With the metrology corner cube conjugated ~36 m behind the telescope pivot points, the differential pointing corresponds to a baseline motion of 0.6 mm. Depending on which baseline direction is considered, 20 m in declination or 100 m in right ascension, this corresponds to fractional errors of 3e-5 and 6e-6 respectively. In the declination direction, which has the highest sensitivity to baseline errors, the estimation above (3e-5 or 0.6 mm) represents a good fraction of the observed baseline errors (7.3e-5 or 1.5 mm).

The analysis above seems to indicate that a strong contributor for the current baseline instability might have been found, which is still the dominating contributor to the astrometric error budget.

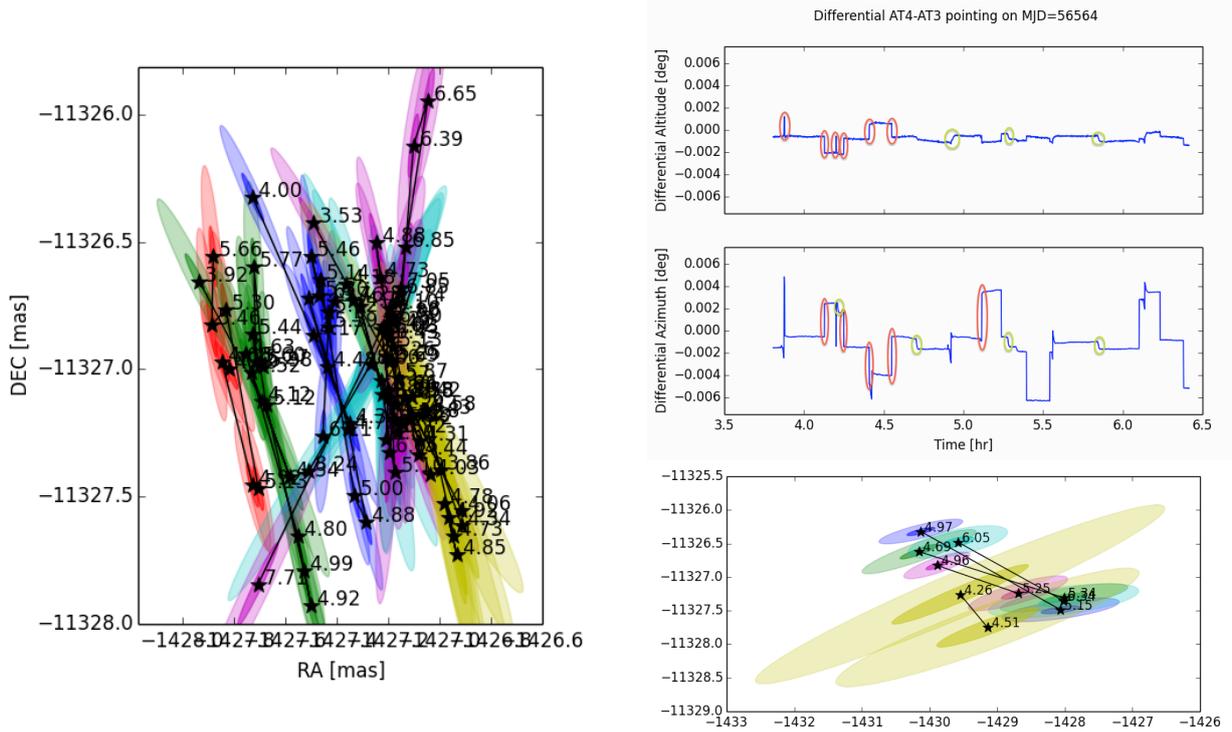

*Figure 5: Left: Evolution of the fitted separation inside each night (one color per night) when limiting the data to a ~1.5 hrs time-sliding interval within a given night. This plot explores the stability of the fit within each night. Upper-Right: For each night (each color), the JW pipeline is used to fit independently the SWAP and NORMAL data. The NORMAL results are in the upper left corner, the SWAPPED in the lower right, highlighting a baseline bias that depends on the SWAP state. Lower-Right: Differential pointing between AT4 and AT3, on MJD=56564, over the observing interval of HD10360. Red circles are recording artifacts, green circles are real differential pointing jumps associated to SWAP <-> NORMAL transitions. Note: the differential azimuth is not corrected from the cosine of the altitude.*

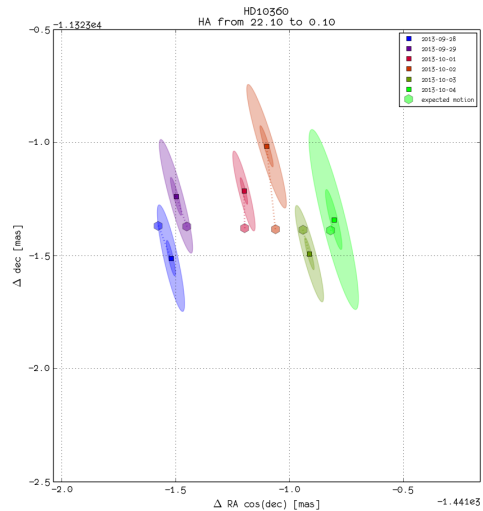

Figure 6: Astrometric fits (AM pipeline) with a modified Chi^2, balancing the weighting between the NORMAL and SWAP states. The residuals are improved to 500 µas peak-to-peak in the DEC direction, within the computed uncertainties.

## 3.3 Performance versus requirements summary

**Primary star limiting magnitude** – Even though fainter primary stars have been observed in the PRIMA configuration, K ~ 5 mag represents a limit for reasonable fringe tracking behavior with limited lock breaks (single fringe tracker locked >80% of the time) in median seeing (1.2 arcsec, 2 ms) and median wind conditions (<10 m/s). Beyond this limit, the degraded fringe tracker performance impacts the open shutter time and the sensitivity on the secondary star. The origin of this performance degradation is a combination of telescope-differential flats and injection fluctuations (details are given in the next section).

**Secondary star sensitivity** – K ~ 9 mag for the secondary star represents a magnitude that can be properly tracked by IRIS the tip/tilt sensor running a very low frame rate. At this magnitude the secondary fringe sensor does not track the fringes but scan through the envelope. With the fringe position measured only a very small fraction of the time, the statistical precision of the measurement drops dramatically, unless compensated by a significant increase of the data collection interval (depending on the actual length of the scan, the efficiency might be affected by a factor x10). The performance degradation of the primary star fringe sensor has an impact on the secondary star sensitivity through a low off-axis long-exposure fringe contrast.

**Astrometric performance** – The short time scale night-to-night stability has been measured to ~800 µas peak-to-peak, in an unfavorable observing configuration: ~10 arcsec North-South pair with the baseline above, projecting to ~30 m North-South. With the somewhat risky assumption that the dominating source of error is a baseline instability that does not scale with the baseline length (this completely ignores the contribution of the differential OPD in the error budget), the performance expected with a 150 meter baseline is ~160 µas. This calculation only gives an indication of what the extension of PRIMA to an orthogonal, ~150 meter long, North-South baseline may provide. In addition, the astrometric performance is given in µas peak-to-peak, rather than µas rms, in order to remind that the underlying errors are not normally distributed. With the evidence that measurements are affected by biases any averaging of the night-to-night errors is not recommended.

**Observing efficiency** – The demonstrated observing efficiency of 2~4 hours reflects the amount of time spent observing the bright pair in the September 2013 run to achieve the astrometric performance of 800 µas peak-to-peak previously mentioned. Both the short time scale (~20 minutes) statistical errors, and the operational efficiency (time spent with shutter open in astrometric collection mode, taking the calibration overhead into account) are compatible with an astrometric data point of 1 hour and a duty cycle (fraction of the time collecting on-sky astrometric data) of 50%.

| Parameter | Demonstrated performance | | Requirements | |
|---|---|---|---|---|
| | Bright case | Individual best | Minimum | Goal |
| Primary star limiting magnitude ($K_{PrS}$) | 3.5 mag | ~5 mag | 7 mag | 8 mag |
| Secondary star sensitivity at 10" ($K_{SeS}$) | 3.5 mag | ~9 mag | 13 mag / 11 mag | 14 mag |
| Astrometric precision and accuracy for 10 arcsec separation (µas) | ~160 µas (extrapolated) short term peak-to-peak stability only | ~160 µas (extrapolated) | 50 µas / 20 µas long term rms stability and accuracy | 20 µas |
| Astrometric data point duration per baseline, assuming 50% open shutter | 2~4 hr | 1 hr | 1 hr | |

Table 2: Top-level summary of the demonstrated performance of PRIMA versus ESPRI exoplanet program requirements. The demonstrated performance presents the bright case outcome of the September 2013 run, and the individual best of each parameter (160 µas on a K ~ 9 mag secondary star in 1 hr was not demonstrated). The requirements, derived from the ESPRI program requirements, show two scenarios: fainter secondary (to observe more targets) or increased astrometric performance (to detect less massive planets). Whereas the astrometric precision and accuracy requirements are given in µas rms, the demonstrated performance is in µas peak-to-peak: this is to convey the idea that the demonstrated performance suffers from strong biases of unknown statistical behavior. Care must be taken when comparing the demonstrated numbers to the requirements.

## 4 FUTURE PLANS

Based on the September 2013 observations, the instability of the narrow-angle astrometric baseline probably stayed the limiting factor of the astrometric performance. However, bringing this baseline under control does not guarantee that the astrometric performance requirement will be met: other error terms might become significant. This section presents the planed improvements and investigations considered to reach the next level of PRIMA Astrometry performance.

### 4.1 Internal metrology extended to primary space

Locating the metrology end-point, a corner cube retro-reflector, at the secondary mirror, the primary space conjugate of the astrometric baseline is located ~35 m behind the telescope pivot. As such, the astrometric baseline vector depends on the location of the corner cube and the conjugation process to primary space. A baseline stability of 150 μm ($10^{-6}$ on a 150 meter long baseline) corresponds to a telescope pointing knowledge of 900 mas, or a retro-reflector transverse position stability of 5 μm. Both numbers seem impossible to meet without an extension of the metrology to primary space, so that no conjugation process gets in the way of the narrow angle baseline realization. For PRIMET to reach primary space, the metrology has to operate with an offset from the telescope optical axis, as the central obscuration of the telescope makes primary space invisible on axis. A pupil plane illustration of the proposed configuration is shown in Figure 7: a retro-reflector is installed on the telescope spider, on the side of the secondary mirror.

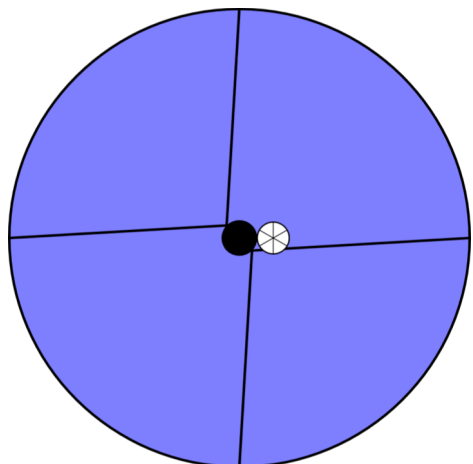

Figure 7: AT pupil of 1.8-meter diameter (blue) with 140 mm secondary mirror central obscuration and spiders (black). A 125 mm diameter retro-reflector is installed on the side of the secondary mirror.

With just this modification, the system is however in a configuration in which there is an offset between the effective center of the stellar pupil and the metrology endpoint. This offset between imaging and astrometric baseline increases the susceptibility to the product of pupil and tilt astrometric errors[6]. The narrow angle baseline is the vector joining the metrology endpoints. For a single mode instrument like the FSU, the imaging baseline is the vector joining the barycenter of the overlap between the telescope pupil and the FSU fiber mode. With the metrology and the FSU co-aligned in tilt and shear, the FSU fiber mode is centered on the metrology corner cube. The barycenter of the overlap is therefore somewhere between the telescope axis (center of the telescope pupil) and the metrology corner cube (center of the FSU fiber mode). The offset between imaging baseline and narrow angle baseline can be approximated to half the off-axis distance of the metrology corner cube from the optical axis. For a 125 mm corner cube next to a 140 mm secondary, this distance is on the order of 132.5 mm. Assuming a tenth of a K-band point spread function tilt error and a millimeter pupil shift, this corresponds to an internal optical path difference error of 16 nm, equivalent to 22 μas. A shift of the imaging baseline onto the astrometric baseline would still be achievable by modifying the telescope pupil. Installing appropriate masks in front of the secondary mirror could do this, at the expense of a slightly lowered transmission.

### 4.2 New baseline metrology

To satisfy the requirement of an exoplanet science case, the narrow-angle baseline vector must be known, with respect to the earth (International Terrestrial Reference Frame, ITRF), with a precision/accuracy of 150 μm. This accuracy on the baseline knowledge must be maintained over the entire timescale of an astrometric campaign: 4~7 years. The long-term stability of the baseline cannot rely solely on the long-term stability of a removable metrology retro-reflector installed on telescopes that are regularly relocated. Therefore, measuring this retro-reflector position with respect to a stable permanent ground reference seems like a minimum requirement.

The proposed concept to continuously measure the position of the retro-reflector is illustrated in Figure 8. A laser tracker per telescope observes in sequence the retro-reflector in primary space next to the secondary mirror and local redundant references anchored in the ground. This system would provide a measurement of the retro-reflector position at each telescope within its own ground reference frame. Then, the ground reference frames need to be positioned (translation and rotation) with respect to each other, and last with respect to the ITRF. With the long-term stability requirement transferred to the ground references, the retro-reflector in primary space does not have to stay permanently mounted on the telescopes, for operations other then PRIMA-Astrometry. An on-sky wide-angle baseline model, corrected from the external baseline metrology would provide the final transformation to the ITRF. Even though the current wide-angle baseline model residuals are on the order of 150 μm, the joint use of PRIMET and the baseline metrology will improve the precision of the wide-angle process. Finally, the use of stable stellar pairs could help with the calibration and measurement of the astrometric baseline, serving as long-term reference.

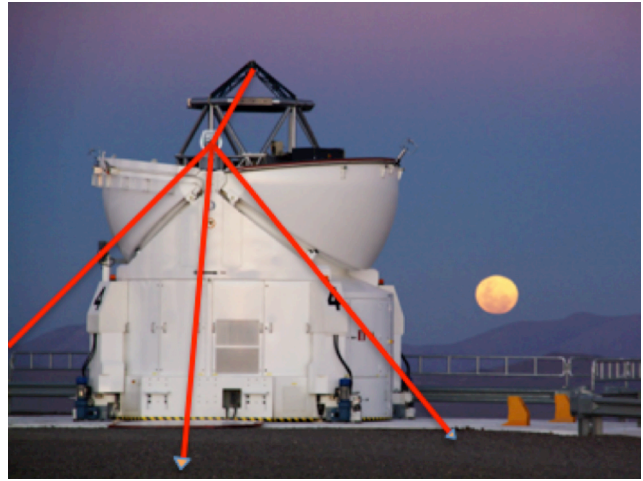

Figure 8: An external baseline metrology concept. Using a 3D laser tracker, the position of the metrology corner cube, installed in primary space next to the secondary mirror, is measured with respect to redundant reference points,

### 4.3 Limiting magnitude of Fast Bright Fringe Tracker

Based on the transmission to the FSU, the fringe-tracking limiting magnitude at 1 kHz, set by a phase standard-deviation of 0.2 rad (SNR=5 on phase estimation), is expected at K=6.9. However, this simplistic limiting magnitude calculation does not represent the typically achieved performance of the FSU. On the September 2013 run, under good conditions, the primary star fringe tracker was satisfactory at K=4.15. For fainter targets, the fraction of time spent with the fringe tracker locked on fringes decreases sharply. In Sahlmann et al. 2009[7], figures 20 and 21 confirm an acceptable tracking behavior up to K=5~6 in single-feed mode, corresponding to K=4~5 in the PRIMA dual-feed mode, followed by a sharp performance degradation. Even in the poorest seeing conditions, the piston should not be able to break the fringe tracker loop. At 1 kHz, the standard deviation of the difference between two consecutive atmospheric pistons is much smaller than pi radians, the phase unwrapping threshold. Instrumental vibrations are not impacting either the stability of the loop: on the ATs they contribute below the atmospheric turbulence. For bright pairs, the fraction of time with the FSU locked on fringe correlates to the amount of flux dropouts seen in the photometric calibration measuremnts. These intensity fluctuations impact the fringe SNR used by the FSU state machine to keep the fringe tracker locked. The SNR threshold under which the FSU drops out of lock is related to an estimation of the SNR fluctuations observed outside the fringe coherence envelope. As pointed out by Sahlmann et al. 2009 (Fig. 19), the fringe SNR outside the fringe envelope is particularly high for the FSU (fringe contrast of 25~30% rms, outside the coherence envelope); it is larger than the photon noise bias. It can be related to a telescope-dependent differential injection into the ABCD fibers. Without photometric channels, only the mean ABCD responses for AT3 and AT4 combined can be corrected (Sahlmann et al. 2009[7], Eq. 2, and section 5.1.2); any photometric fluctuation around the means generates a false fringe signal, responsible for this high SNR.

To improve the behavior of the fringe tracker and reach the limiting magnitude required by the exoplanet program (K~7) the injection stability into the fringe tracker must be improved and the issue of the telescope-differential flat solved. The injection stability is directly related to the stability of the wavefront delivered by the Auxiliary Telescopes. At the moment, only the Tip/Tilt mode is corrected with STRAP in closed loop with M6, and IRIS in closed loop with the STS field steering mirrors, providing a performance of 50 mas rms on bright targets. The injection fluctuations are therefore dominated by the uncorrected higher order modes of the atmospheric turbulence (48% Strehl at 2.2 μm for 0.8 arcsec seeing, per the VLTI ICD, section 3.2.8.2). To reduce the injection fluctuation an adaptive optic system must be installed on the ATs: this is the NAOMI project[8].

### 4.4 Sensitivity of Slow Faint Fringe Sensor

With the secondary fringe tracker slowing down, the thermal background becomes an important factor in the estimation of the phase. For a K=14 magnitude star, and an integration period of 30 minutes for an astrometric data point, this contribution corresponds to an OPD error of 3.5 nm, equivalent to an astrometric error of 5 µas. In order to be background limited, $N_{bkg} > 10\ RON^2 = 46820\ e^{-2}$, an integration period longer than ~1 second is needed. For indication, at an integration period of 1 second, the $SNR_\Phi$ on the phase estimation is 2.5 (~0.4 rad).

At the moment, the FSU is not completely operational below 1kHz. This is more a consequence of the emphasis on bright astrometry than a fundamental limitation. There are issues with the noise properties of the FSU as the operating frequency decreases. It seems to be a combination of pick-up noise, and potentially errors in the processing of the Non-Destructive-Readouts (as seen in the noisier and negative background values at frequencies lower than 1 kHz). At this point, a careful investigation of the slow fringe rate behavior of the FSU is needed.

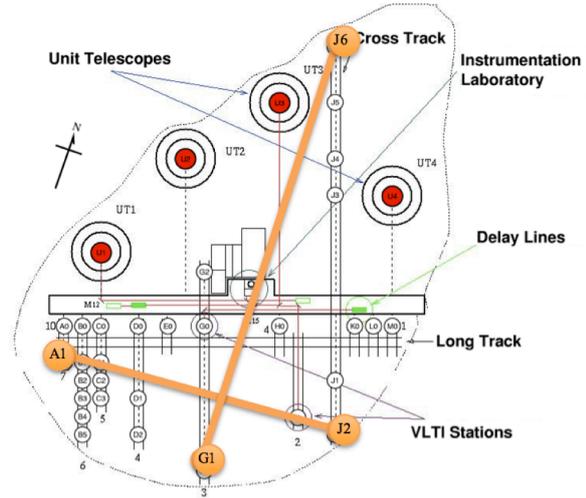

Figure 9: The PRIMA baselines. AT3@A1 - AT4@J2 is the current east-west baseline. AT1@G1 - AT2@J6 is a candidate for the next north-south baseline.

### 4.5 Internal Fringes: Fringe Sensor vs Metrology

At the moment, this is the aspect of the astrometric error budget with the most unknowns. Individually, FSU and PRIMET are relatively well understood. However, whether they provide in combination a sufficient astrometric capability needs to be demonstrated. The absence of significant bias induced by the swap operation needs to be confirmed.

With the baseline errors dominating the error budget, the error terms mentioned in the previous section are difficult to quantify. An auto-collimation mode must be implemented, representative of the on-sky observing conditions, excluding the baseline and atmospheric contributions, therefore limited to the assessment of the differential OPD. Equipment exists to carry out auto-collimation tests including the main delay lines: the light of the calibration source MARCEL is sent through the DDL, toward retro-reflector manually installed on the telescope side of the main delay-lines, to be sent back to the FSU. The drawbacks of this setup are a limitation to a half pupil for the K-band beam, and no inclusion of the STS and telescope optics. Auto-collimation has also been demonstrated to the Nasmyth focus of the ATs, therefore including the STS, derotator, and azimuth axis, with the same MARCEL source. An ideal auto-collimation setup would include M3 (the last mirror with polarization impacts) with a retro-reflection point between M2 and M3, and have full pupil light sources injected before the FSU beam-splitter. Beam-walk, air dispersion, source temperature dependence, long-term stability are example of effects that could be independently verified in the comfort of a controlled environment. The astrometric performance of the instrument without the baseline could be measured directly, carrying astrometric sequences in auto-collimation.

### 4.6 North-South baseline

Until now, the PRIMA project focused on demonstrating astrometry on a single A1-J2 baseline, between AT3 and AT4, the only two ATs ready to host star separators. This east-west baseline provides astrometric measurement in a single orientation. Opening a north-south baseline is part of the PRIMA plan to provide astrometric data points in the orthogonal direction. The current and planned baselines are illustrated in Figure 9. Even though the local compensation of the derotator retardance seemed satisfactory, from the FSU perspective, this was confirmed only on a symmetric South-South baseline. When the North-South baseline is open, the effect of residual differential polarization, between the two rotators operating at 90 deg offset of each other, will have to be considered.

## 5  CONCLUSION

The purpose of the PRIMA review held in January 2014 was to confirm the understanding of the technical challenges facing PRIMA-Astrometry and the credibility of the moving forward plan. The mostly external board of this review was also asked to assess the ESPRI science case in view of the rapid progress in the field of exoplanet science and the delays of PRIMA. This analysis was particularly critical in the context of the launching of the Gaia spacecraft[9] at the end of 2013. Gaia will carry out precision astrometry from space and will massively reduce the areas in which PRIMA could make new discoveries.

The review confirmed that the recovery plan was sound: the core goal of the plan, to control the baseline instability, was well justified. It also confirmed the assessment that the risks are not small, and the effort/time needed to carry out the recovery plan may be significantly underestimated. The recovery plan, presented in Section 4, needed key ESO staff to work essentially full time on PRIMA for at least three years. Taking the high risks involved into account, it is likely that this is more realistically 3-5 years. In terms of science, the accumulated delay and projected delay to completion mean that realistically the ESPRI consortium cannot carry out their planned program in a timely way. This is not the same as saying that PRIMA science is no longer relevant – the board identified a significant potential for astrometric follow-up of Gaia from the ground, but this was not the ESPRI science case.

Based on this, and recognizing that the same key staff are also critical to the preparation of the VLTI for Gravity and Matisse in the same time frame, ESO decided to take a proposal to the April STC that PRIMA be cancelled, and that ESO resources be concentrated on ensuring that Gravity and Matisse are a success. This proposal was recommended by the STC in May 2014, and endorsed by ESO.

___o☺o___